\documentclass[aps,prl,twocolumn,notitlepage,showpacs,floatfix,superscriptaddress]{revtex4-1}
\usepackage{dcolumn}
\usepackage{tabularx}
\usepackage{bm}
\usepackage{soul}
\usepackage{amsmath,amssymb,graphicx}
\usepackage[colorlinks=true,citecolor=blue,urlcolor=blue,linkcolor=blue]{hyperref}
\usepackage{environ}

\NewEnviron{eqnsplit}{%
\begin{equation}
\begin{split}
  \BODY
\end{split}
\end{equation}
}

\newcommand{\mrm}[1]{\mathrm{#1}}

\begin{document}
\title{Sensing chiral magnetic noise via quantum impurity relaxometry}

\author{Avinash Rustagi}
\email{arustag@purdue.edu}
\affiliation{School of Electrical and Computer Engineering, Purdue University, West Lafayette, IN 47907}

\author{Iacopo Bertelli}
\affiliation{Department of Quantum Nanoscience, Kavli Institute of Nanoscience, Delft University of Technology, Lorentzweg 1, 2628 CJ, Delft, The Netherlands}
\affiliation{Huygens--Kamerlingh Onnes Laboratorium, Leiden University, Niels Bohrweg 2, 2300 RA, Leiden, The Netherlands}

\author{Toeno van der Sar}
\affiliation{Department of Quantum Nanoscience, Kavli Institute of Nanoscience, Delft University of Technology, Lorentzweg 1, 2628 CJ, Delft, The Netherlands}

\author{Pramey Upadhyaya}
\email{prameyup@purdue.edu}
\affiliation{School of Electrical and Computer Engineering, Purdue University, West Lafayette, IN 47907}
\date{\today}
\begin{abstract}
We present a theory for quantum impurity relaxometry of magnons in thin films, exhibiting quantitative agreement with recent experiments without needing arbitrary scale factors used in theoretical models thus far. Our theory reveals that chiral coupling between prototypical spin$>$1/2 quantum impurities and magnons plays a central role in determining impurity relaxation, which is further corroborated by our experiments on nickel films interfaced with nitrogen-vacancy centers. Along with advancing magnonics and understanding decoherence in hybrid quantum platforms with magnets, the ability of a quantum impurity spin to sense chiral magnetic noise presents an opportunity to probe chiral phenomena in condensed matter.
\end{abstract}
\maketitle
\textit{Introduction}| Magnons -- quanta of spin wave excitations -- are fundamental to the understanding of the dynamical properties of magnetically ordered materials. This understanding forms the basis for creating next-generation classical and hybrid quantum technologies in magnonics (an emerging field utilizing magnons as information carriers) \citep{chumak2015magnon,li2020hybrid,lachance2019hybrid}, potentially enabling magnon-mediated coherent control \cite{andrich2017long} and coupling of distant quantum spins \cite{trifunovic2013long}. In addition, topological qubit platforms, typically, involve magnetic materials \cite{alicea2012new} which could introduce an additional source of decoherence. The rapidly growing field of quantum technology involving magnetic materials makes it imperative to understand the decoherence introduced in quantum systems placed in close proximity to magnetic materials. As such, it is critical to develop sensitive novel probes for studying the dynamical properties of magnetically ordered materials.

Quantum impurity (QI) relaxometry \citep{QSensing_RMP} -- a sensing scheme measuring the relaxation rate of an impurity spin due to its coupling with magnetic noise \citep{ariyaratne2018nanoscale,casola2018probing,Demler2019noiseMagnetometry} -- has recently emerged as a sensitive, local and non-invasive technique for probing condensed matter systems including magnetic materials \cite{van2015nanometre,purser2020spinwave,lee2020nanoscale,du2017control,Flebus2018relaxometry,Flebus2018_AFM,finco2020AFMimaging}. QIs coupled to magnetic thin films form model systems for developing an understanding of decoherence introduced in qubits that are in close proximity to magnetic materials. It is, therefore, important to develop a predictive model for understanding QI-relaxometry of thin film magnons.
\begin{figure}[hbtp]
\centering
\includegraphics[width=0.49\textwidth]{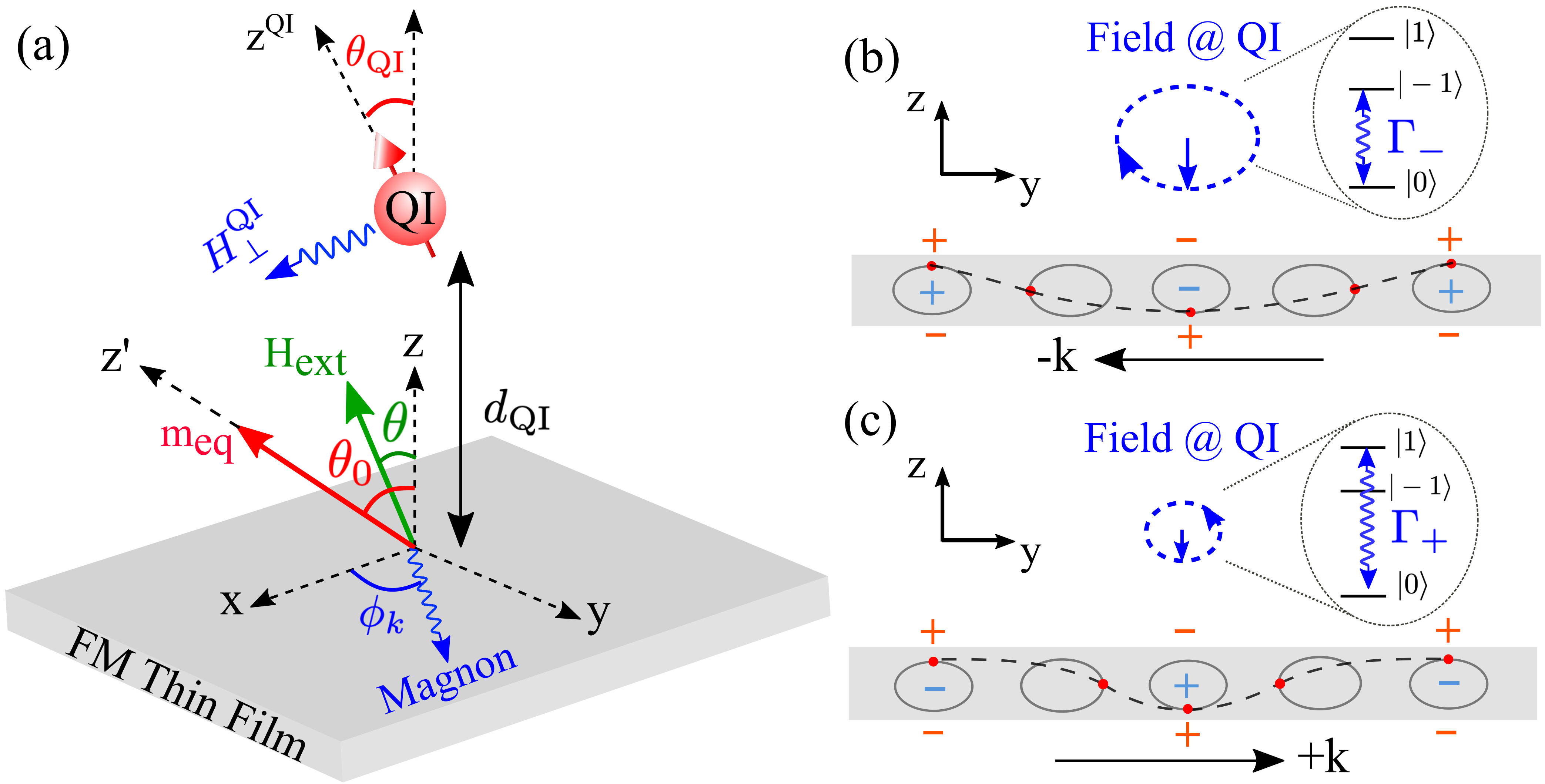}
\caption{\label{Schematic}(a) Schematic of a QI-FM hybrid where the external field is applied along the QI quantization axis ($\theta_{QI}=\theta$) and the direction of equilibrium magnetization is along $\hat{z}^\prime$ (minima of the magnets free energy). Oppositely handed chiral fields emanating from (b) left and (c) right moving magnons couple to (0,-1) and (0,+1) QI transitions, respectively \citep{ellipticfield}. The effective volume (in lightblue color) and surface (in orange color) magnetic charges add (subtract) for left (right) moving magnons generating stronger (weaker) dipolar field at the QI.}
\end{figure}

Chirality plays a central role for coupling magnons in thin films to QIs. On the one hand, spin transitions are only driven efficiently by rotating magnetic fields of the correct handedness. On the other hand, magnons produce magnetic fields with a handedness that depends on their travel direction (i.e., their fields are chiral).

The zero-field splitting (ZFS) present in typical spin$>$1/2 QIs, such as nitrogen-vacancy (NV) centers in diamond \citep{lee2017topicalNV,hopper2018spin} and silicon-vacancy (SiV) centers in SiC \cite{kraus2014room}, gives rise to opposite effective built-in fields causing the different electron spin resonance (ESR) transitions to respond to magnetic field of opposite handedness (provided the applied field is smaller than the ZFS) as demonstrated in NV-centers \citep{Alegre2007ODMR}. Additionally, counter-propagating magnons with non-zero wavevector component transverse to the equilibrium magnetization and finite out-of-plane deviation (such as, the Damon-Eschbach modes of thin magnetic films \citep{damon1961magnetostatic}) produce counter-rotating magnetic fields having \textit{unequal} amplitudes \cite{tao2019chiral}. This results from the combined effect of bulk $\rho_m \propto - \vec{\nabla}\cdot\vec{m}$ and surface $\sigma_m \propto \vec{m}\cdot\hat{n}$ (where $\hat{n}$ is the surface normal) magnetic charges (cf. Fig.~\ref{Schematic}b-c). Consequently, the $m_s= 0 \rightarrow -1$ and $m_s=0\rightarrow +1$ transitions of a prototypical spin-1 QI are driven by magnon-generated fields of different magnitudes. Theoretical models used to analyze experiments \citep{van2015nanometre,du2017control,purser2020spinwave}, by excluding out-of-plane magnetization fluctuations (and thus $\sigma_m$), neglect the role of chirality and require arbitrary scale factors of unknown origin to quantitatively fit the experimentally measured relaxation rates for the $m_s= 0 \rightarrow -1$ and $m_s=0\rightarrow +1$ transitions.

In this Letter, by combining the general theoretical framework of quantum relaxometry \citep{Flebus2018relaxometry,Demler2019noiseMagnetometry} with Landau-Lifshitz-Gilbert (LLG) phenomenology \cite{landau1980statistical,*gilbert2004phenomenological} for magnon dynamics in thin magnetic films, we construct a theory for QI-relaxometry of magnons which inherently captures the chiral coupling. As central results, we show that our theory (i) results in excellent quantitative agreement with recent experiments without introducing arbitrary scale factors, and (ii) predicts crossover between relaxation rate for the $m_s= 0 \rightarrow -1$ and $m_s=0\rightarrow +1$ transitions as a function of applied field as an experimental signature of the chiral coupling. We corroborate the latter by presenting new data on Nickel thin film interfaced with NV center QI. Our results highlight the importance of chirality in constructing predictive models for advancing magnonics via QI-relaxometry. More generally, they suggest that: (i) chirality of magnon-generated fields is essential in governing decoherence of quantum systems proximal to magnetic materials, and,  (ii) QI-relaxometry can be extended to non-invasively and locally probe the physics of chiral electronic \citep{hasan2010rmp,tokura2019magnetic,li2019intrinsic} and magnetic modes \citep{shindou2013chiral,kim2017chiral} living in condensed matter systems of interest via the magnetic noise emanating from them.

\textit{Relaxation Model}| The hybrid includes a ferromagnetic thin film of thickness $L$ and a QI located at a height $d_{QI}$ above the thin film (cf. Fig.~\ref{Schematic}a). To evaluate the QI relaxation dynamics arising from coupling to the magnetic noise emanating from the film at room temperature, we begin by recasting the relaxation rates of QIs \citep{Flebus2018relaxometry,Demler2019noiseMagnetometry} in a classical form relating them to the correlations between magnetization deviations. Next, we evaluate these correlations within the framework of linearized-LLG.

Here, we consider a prototypical spin-1 QI (like NV center) with spin-triplet ground state $\vert m_s \rangle$ labeled by the projections $m_s = \{-1,0,1\}$ along the QI quantization axis. In the presence of a magnetic field $\vec{H}^{QI}$, the Hamiltonian of effective two level systems (TLSs) formed by the states ($\vert +1 \rangle, \vert 0 \rangle$) and ($\vert 0 \rangle, \vert -1 \rangle$) denoted by $\xi_+$ and $\xi_-$, respectively, can be written as 
\begin{equation}
\label{NVHamiltonian}
\begin{split}
\mathcal{H}^{\xi_\pm}_{TLS} &=\dfrac{\omega_\pm}{2} [\mathcal{I} \pm \sigma_z] + \dfrac{\gamma}{2\sqrt{2}} [H_+^{QI} \sigma_- + H_-^{QI} \sigma_+], \\
\end{split}
\end{equation}
where $\gamma$ is the gyromagnetic ratio, $\omega_\pm^{QI} = \Delta \pm \gamma H_z^{QI}$ are the QI-ESR frequencies with the ZFS $\Delta$,  and $H_\pm^{QI} = H_x^{QI} \pm i H_y^{QI}$. The superscript `QI' denotes that the field components are evaluated in the QI-frame (where the z-axis is aligned along $z^\mrm{QI}$) attained via the rotation matrix $R_{yz}(\theta_{QI},\phi_{QI})$ (representing rotation about the z-axis by $\phi_{QI}$ followed by a rotation about the y-axis by $\theta_{QI}$). This form of the Hamiltonian confirms that $m_s= 0 \rightarrow -1$ and $m_s=0\rightarrow +1$ transitions are caused by fields of opposite handedness $H_+^{QI}$ and $H_-^{QI}$, respectively.

The rates corresponding to the transitions $\vert 0 \rangle \rightarrow \vert \mp 1 \rangle$ (marked by subscript $\mp$) are given by the spectral density of the field perpendicular to the quantization axis evaluated at the ESR frequencies $\omega_\mp$ \citep{Flebus2018relaxometry,Demler2019noiseMagnetometry}: $\Gamma_{\mp}(\omega_{\mp}) = ({\gamma}^2/2) \int dt \, e^{i\omega_\mp t} \big\langle H^{QI}_\pm (t) \, H^{QI}_\mp (0) \big\rangle $. Here, $\langle ... \rangle$ denotes averaging over the noise realizations.  

The Fourier component of the field at QI, due to a spin wave mode of the film (with an amplitude $\delta\vec{m}^\prime(\vec{k})$, frequency $\omega$, and wavevector $\vec{k}$) can be written as:
 $\vec{H}^{QI}(\vec{k}) = \mathcal{D}^\mathrm{eff} (\vec{k})\delta\vec{m}^\prime(\vec{k})$.  
Here, $\delta\vec{m}^\prime(\vec{k})$ is the magnetization deviation in the frame where the z-axis is aligned along $\vec{m}_{eq}$ (magnet frame), and $\mathcal{D}^\mathrm{eff} (\vec{k}) = R_{yz}(\theta_{QI},\phi_{QI}) \mathcal{D}(\vec{k}) R_y^T(\theta_0)$ is the `rotated' dipolar tensor, given
\begin{equation}
\label{dipolarmatrix}
\begin{split}
\mathcal{D}(\vec{k}) &= -2\pi A_k \left( \begin{array}{ccc}
\cos^2 \phi_k & \sin2\phi_k/2 & i\, \cos\phi_k \\ 
\sin2\phi_k/2 & \sin^2\phi_k & i\, \sin\phi_k \\ 
i\, \cos\phi_k  & i\, \sin\phi_k  & -1
\end{array} \right),
\end{split}
\end{equation}
with $A_k= M_s  \, e^{-k d_{QI}} \left[1-e^{-kL}\right]$ \cite{guslienko2011magnetostatic}. Typically, $H_\mrm{ext}$ is aligned with the QI axis (see Fig.~\ref{Schematic}a), thus $\theta_{QI}=\theta$ and we choose $\phi_{QI}=0$.

Substituting the fields from spin waves into the relaxation rate equation of QI, we get 
\begin{equation}
\label{rate_exp}
\begin{split}
\Gamma_{\mp}(\omega_\mp) \!= \!\dfrac{{\gamma}^2}{2} \!\int \! \dfrac{d\vec{k}}{(2\pi)^2}\!\!\! \sum_{i,j \in \{x,y\}} \!\!\mathcal{D}^\mathrm{eff}_{\pm i} (\vec{k}) \mathcal{D}^\mathrm{eff}_{\mp j} (-\vec{k}) C_{ij}(\omega_\mp,\vec{k}),
\end{split}
\end{equation}
where $\mathcal{D}^\mathrm{eff}_{\pm \nu} = \mathcal{D}^\mathrm{eff}_{x \nu } \pm i \mathcal{D}^\mathrm{eff}_{y \nu}$, and the correlations between the magnetization deviations $C_{ij}(\omega,\vec{k})$ is the Fourier transform of $C_{ij}(\vec{r}-\vec{r}\,^\prime, t-t^\prime) = \langle \delta m_i^\prime (\vec{r}, t) \, \delta m_j^\prime (\vec{r}\,^\prime, t^\prime)\rangle$. Eq.~(\ref{rate_exp}) relates the QI relaxation to the magnetization correlations, demonstrating the importance of both auto- and cross-correlations in the QI spin's dynamics, which we evaluate next for thermally populated magnons.

\textit{Magnon correlations}| The magnetization dynamics is governed by the LLG equation $\dot{\vec{m}} = -\gamma\, \vec{m} \times \left[\vec{H}_{\text{eff}} + \vec{h} \right] + \alpha \, \vec{m} \times \dot{\vec{m}}$, where $\alpha$ is the Gilbert damping, $\vec{h}$ is the excitation field, and effective field $\vec{H}_{\text{eff}} = -\partial_{\vec{m}} (\mathcal{F}/M_s)$. Here, the free energy density describing the ferromagnetic thin film includes the Zeeman, exchange, and dipole-dipole energy terms and is given by $\mathcal{F} = -M_s \left(\vec{H}_\mrm{ext} + \vec{H}_D/2 \right) \cdot \vec{m} + A_{ex} \sum_{i\in \{x,y\}} (\partial_i \vec{m})^2$ where $\vec{H}_\mrm{ext}$ is the external field, $A_{ex}$ is the exchange constant, $M_s$ is the saturation magnetization, $\vec{m}$ is the magnetization unit vector, and $\vec{H}_D$ is the demagnetization field \cite{demagfield}. We find the magnetic susceptibility in response to $\vec{h}^\prime$ (in the magnet frame) by solving the linearized-LLG in Fourier domain about the equilibrium magnetization giving $\delta m_i^\prime(\omega,\vec{k}) = S_{ij}(\omega,\vec{k}) h_j^\prime(\omega,\vec{k})$. The susceptibility matrix is given by
\begin{equation}
\label{sus_matrix}
S(\omega,\vec{k}) = \dfrac{\gamma}{\Lambda} \left( \begin{array}{cc}
\omega_3 - i \alpha \omega & -\omega_1 - i \omega \\ 
-\omega_1 + i \omega & \omega_2 - i \alpha \omega 
\end{array} \right),
\end{equation}
where
\begin{equation}
\begin{split}
\omega_1 &= \gamma H_d \sin\phi_k \cos\phi_k \cos\theta_0 \\
\omega_2 &= \omega_0 + \gamma H_d [f_k \cos^2\phi_k \cos^2\theta_0 + (1-f_k) \sin^2\theta_0]\\
\omega_3 &= \omega_0 + \gamma H_d f_k \sin^2\phi_k \\
\omega_0 &= \gamma [H_\mrm{ext} \cos(\theta_0-\theta) - H_d \cos^2\theta_0 + H_{ex} k^2] \\
\Lambda &= (\omega_2-i\alpha\omega)(\omega_3-i\alpha\omega) - \omega_1^2 - \omega^2.
\end{split}
\end{equation}
with $H_d = 4\pi M_s$ and $f_k = 1 - (1-\exp(-k L))/(kL)$. The spin-wave dispersion evaluated from the pole of the susceptibility is given by \citep{kalinikos1986theory} $\omega_{sw}^2 (\vec{k}) = \omega_2 \omega_3 - \omega_1^2$.

In thermal equilibrium, $\vec{h}^\prime$ is the thermal stochastic field with zero mean and local, instantaneous correlation \citep{brown1963thermal,kubo1970brownian}
$\langle h_i^\prime(\vec{r},t)\, h_j^\prime(\vec{r}\,^\prime,t') \rangle = 2 D_{th} \delta_{ij} \delta(t-t')\delta(\vec{r}-\vec{r}\,^\prime)$ where $D_{th}=\alpha k_B T/(\gamma M_s L)$. Using these stochastic field correlations and the susceptibility matrix in Eq.~(\ref{sus_matrix}), we can determine the magnetization correlations in Fourier space
\begin{equation}
\label{corr_functions}
\begin{split}
C_{ij}(\omega,\vec{k}) &= 2 D_{th} \sum_{\nu = \{ x,y\}} S_{i \nu} (\omega,\vec{k}) S_{j \nu} (-\omega,-\vec{k}).
\end{split}
\end{equation}

Eqs.~(\ref{corr_functions}) and (\ref{rate_exp}) are the central theoretical results of our work, describing the impact of fluctuating dynamics of magnons in thin films on the QI-spin relaxation. Particularly, we highlight that as per Eq.~(\ref{corr_functions}), all the correlators $C_{xx}$, $C_{xy}$, $C_{yx}$, and $C_{yy}$ are non-zero. This amounts to including both out-of-plane and in-plane magnetization deviations arising from finite ellipticity of magnons, thereby including the effect of both surface and bulk magnetic charges. Our theory thus inherently captures the chiral nature of magnon noise resulting from the combination of surface and bulk magnetic charges [c.f {Fig.~\ref{Schematic}b-c}], and its impact on relaxation rates via Eq.~(\ref{rate_exp}). We thus refer to our theory as `chiral' theory in the following. Taking the limit where only the in-plane magnetization deviations are included (equivalent to setting $C_{xx} = C_{xy} = C_{yx}=0$) and modeling the non-zero magnetization correlation ($C_{yy}$) by a Lorentzian, Eqs.~(\ref{rate_exp}) and (\ref{corr_functions}) reduces to the existing theoretical models used in Refs\cite{van2015nanometre,du2017control,purser2020spinwave}. Since such models do not include surface charges, the chiral nature of the magnon noise is neglected and we refer to them as the `achiral' theory.
%
%
%
\begin{figure}[htbp]
\centering
\includegraphics[width=0.49\textwidth]{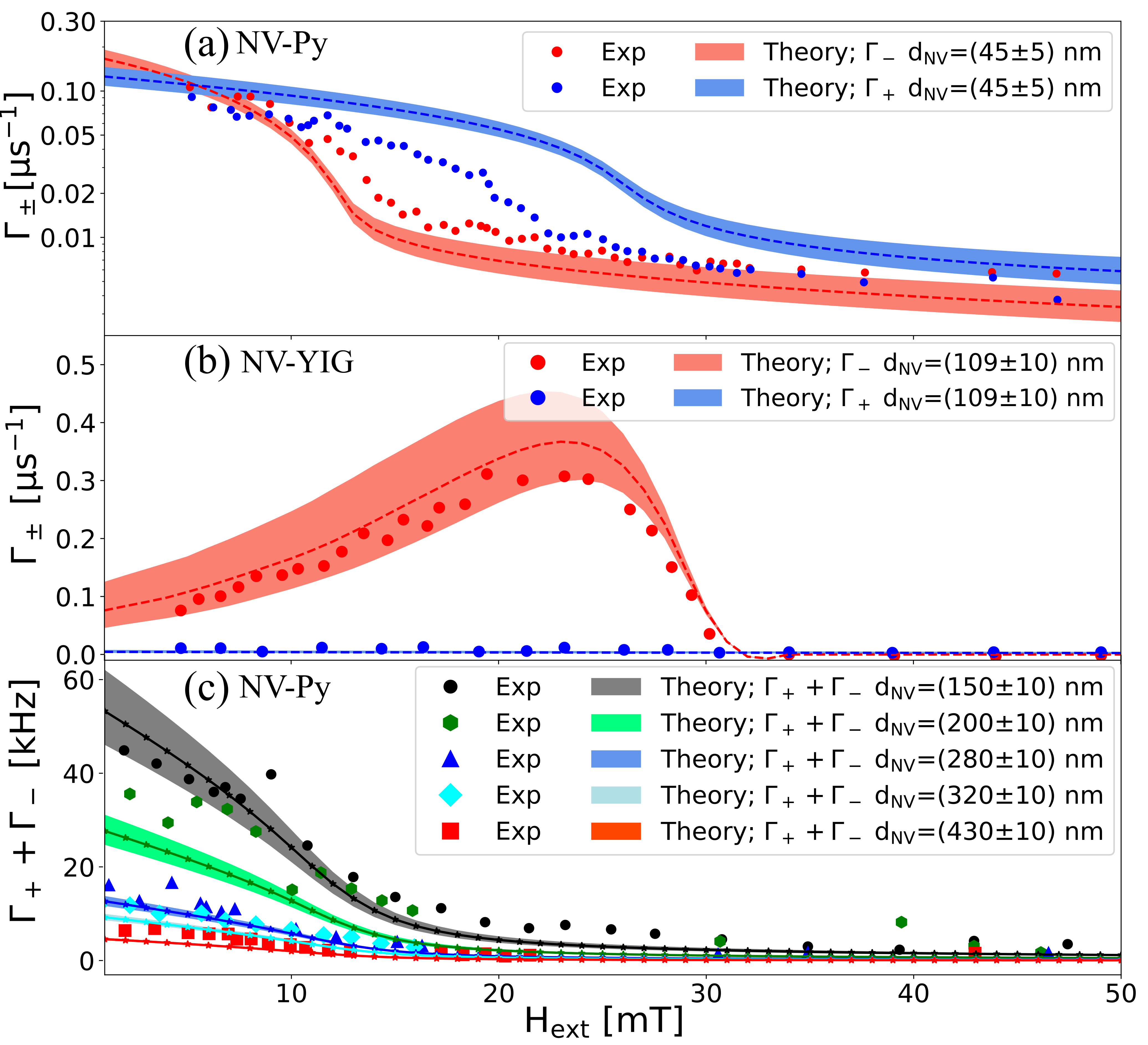}
\caption{\label{Comparison_Science}Comparison of rates evaluated using Eq.~(\ref{rate_exp}) to the experimental measurement in hybrids of NV center - (a) Py ($L$=30 nm, $M_s$=800 emu/cc, $A_{ex}=10^{-6}$ erg/cm, $\alpha=0.015$) \cite{van2015nanometre}, (b) YIG ($L$=20 nm, $M_s$=124 emu/cc, $A_{ex}=3.7 \times 10^{-7}$ erg/cm, $\alpha=0.0001$) \cite{du2017control}, and (c) Py ($L$=20 nm, $M_s$=800 emu/cc, $A_{ex}=10^{-6}$ erg/cm, $\alpha=0.015$) \cite{purser2020spinwave}. The shaded regions correspond to the bounds set by lower and higher NV height, while the dashed lines correspond to the average NV height. The experimental data in -- (a) is adapted from Ref.\cite{van2015nanometre} under the CC BY 4.0 license in Nature Communications (Springer Nature), Copyright 2015, (b) is adapted with permission from Ref.\cite{du2017control}, AAAS, and (c) is adapted from Ref.\cite{purser2020spinwave} with permission from AIP Publishing, Copyright 2020. }
\end{figure}
%
%

\textit{Quantitative Benchmarking}| We begin by bench-marking our chiral theory against recent magnon-relaxometry experiments. Specifically, we consider the hybrid of NV-center QI with (i)  permalloy (Py) \cite{van2015nanometre}(cf. Fig.~\ref{Comparison_Science}a), (ii)  yttrium iron garnet (YIG) \cite{du2017control} (cf. Fig.~\ref{Comparison_Science}b), and (iii) Py,  where, additionally, the NV center-thin film distance was varied \cite{purser2020spinwave} (cf. Fig.~\ref{Comparison_Science}c). In the first two experiments both $\Gamma_+$ and $\Gamma_-$ were measured as a function of $H_\mrm{ext}$, while the third experiment measured the combined rate $\Gamma_+ + \Gamma_-$. To analyze these experiments previous theoretical models used the achiral theory, which required arbitrary scaling factors as  free  parameters  to quantitatively fit the data \cite{du2017control,purser2020spinwave}. Here, we instead apply the chiral theory (using material and geometry parameters from  the  respective  experimental references and mentioned in the caption) with no additional scaling factor. As the first central result of our work, we can see from Fig.~\ref{Comparison_Science} that the theory is in good quantitative agreement with the experimental data requiring no free parameters, validating the presented theoretical formalism.

%
\begin{figure}[htbp]
\centering
\includegraphics[width=0.49\textwidth]{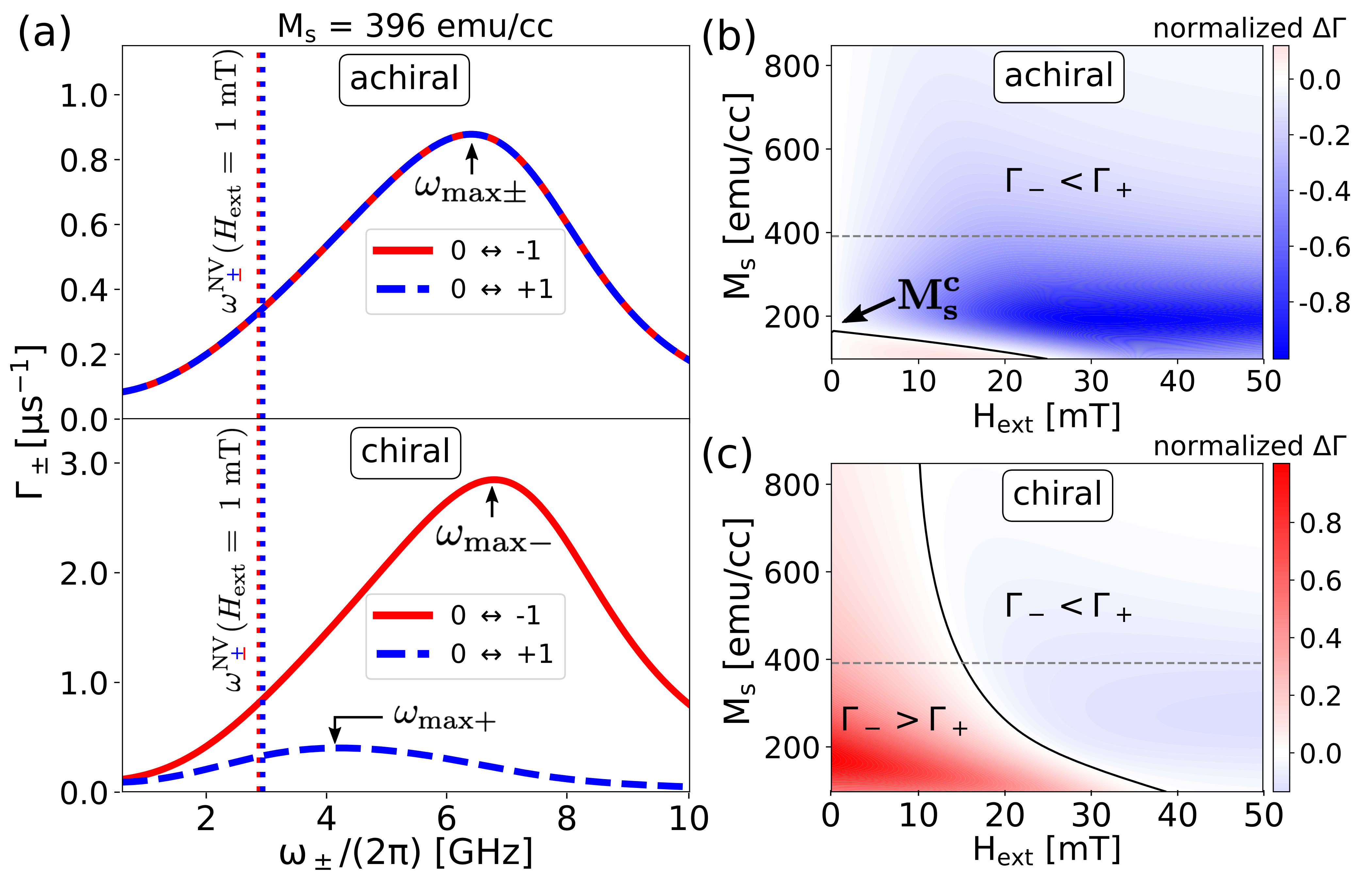}
\caption{\label{TheoryDiff} (a) Computed $\Gamma_\pm$ as a function of $\omega_\pm$ for $M_s$ = 396 emu/cc and $H_\mrm{ext}$=1 mT using the achiral and chiral theory in the top and bottom panels, respectively. The vertical dotted lines mark the NV-ESR frequencies at $H_\mrm{ext}=$1 mT. Normalized NV-ESR transition rate difference $\Delta \Gamma = \Gamma_- - \Gamma_+$ evaluated as a function of $H_\mrm{ext}$ for different magnetic materials parameterized by $M_s$ using the (b) achiral and (c) chiral theory. Parameters used are $L=40$ nm, $\Delta/(2\pi) = 2.87$ GHz, $d_{NV}=40$ nm, $A_{ex} = 8.47\times 10^{-7}$ erg/cm, $\alpha=0.05$, and $\theta_{NV}=\theta=54.75^\circ$.}
\end{figure}
%
\textit{Chiral coupling dependent relaxation}| Having established quantitative agreement between our theory and experiments, we next turn to understand the specific role of chirality. We begin by plotting the relaxation rates $\Gamma_-$ and $\Gamma_+$ of a hypothetical QI spin, whose $\omega_-$ and $\omega_+$ transition frequencies are  scanned  through  the  magnon spectrum.  As highlighted in the introduction, due to the chiral coupling between QI and thin film magnons, $m_s=0\rightarrow+1$ and $m_s=0\rightarrow-1$ transitions are driven by fields of unequal magnitude (c.f Fig.~\ref{Schematic}b-c). Indeed, we observe that the chiral theory consequently predicts $\Gamma_- \neq \Gamma_+$ for $\omega_- = \omega_+$ (cf. Fig.~\ref{TheoryDiff}a--bottom panel). On the other hand, the achiral theory predicts $\Gamma_- = \Gamma_+$ for $\omega_- = \omega_+$ (cf. Fig.~\ref{TheoryDiff}a--top panel). 

While Fig.~\ref{TheoryDiff}a highlights the key difference between the chiral and achiral theory, the relaxometry experiments are performed in the presence of an external magnetic field, which by making $\omega_+$ and $\omega_-$ different, does not allow to directly measure the curve shown in Fig.~\ref{TheoryDiff}a. To look for an experimental signature of the chiral coupling-induced non degenerate relaxation rates, we plot the (normalized) NV-ESR transition  rate  difference $\Delta\Gamma = \Gamma_- - \Gamma_+$ as a function of external field $H_\mrm{ext}$ and for different magnetic materials (as parameterized by their saturation magnetization $M_s$) using the chiral and achiral theory in Fig.~\ref{TheoryDiff}b-c. Within both theories, for $M_s$ below a critical value (referred here as $M_s^c$, which is  equal to $\sim $150  emu/cc for the geometrical parameters of the NV-magnet hybrid as mentioned in the caption of Fig.~\ref{TheoryDiff}), $\Delta\Gamma$ changes sign as $H_\mrm{ext}$ is increased.  In contrast, for $M_s > M_s^c$, only Eq.~(\ref{rate_exp}) predicts a sign change in $\Delta\Gamma$ with $H_\mrm{ext}$. As explained next, the latter is an experimental signature of chiral coupling-induced non-degenerate relaxation rates highlighted in Fig.~\ref{TheoryDiff}a.

To understand the relaxation rate of NV-QI, we add on Fig.~\ref{TheoryDiff}a the location of NV-ESR transitions as dotted vertical lines for $H_\mrm{ext} \approx 0$. We find that $M_s^c$ corresponds to that value of saturation magnetization above which the NVs $\omega_\pm^{NV}$ transitions lie below $\omega_\mrm{max \pm}$ -- the frequency location that would maximize the rates $\Gamma_\pm$ for NV-QI within the achiral theory \cite{theorydiff} (see Fig.~\ref{TheoryDiff}a--top panel). As the field is increased, $\omega_-^{NV}$ ($\omega_+^{NV}$) decreases (increases) linearly (see below Eq.~(\ref{NVHamiltonian})), while $\omega_\mrm{max \pm}$ shifts to a higher frequency as $\gamma \sqrt{H_\mrm{ext}(H_\mrm{ext} + H_d)}$ (as per the shift of magnon bands given by the Kittel formula \cite{kittel1948fmr}). Consequently, with increasing $H_\mrm{ext}$ (for the experimentally relevant range of $H_\mrm{ext} \ll H_d$) $\Gamma_-$ decreases faster than $\Gamma_+$, corresponding to $\omega_-^{NV}$ moving further below $\omega_\mrm{max \pm}$ when compared to $\omega_+^{NV}$. $\Delta\Gamma$, therefore, decreases with increasing $H_\mrm{ext}$ for $M_s > M_s^c$. The key point is that, since in the achiral theory $\Delta\Gamma=0$ for $H_\mrm{ext}= 0$ (as $\omega_+^{NV}=\omega_-^{NV}$ for
$H_\mrm{ext}= 0$; cf. Fig.~\ref{TheoryDiff}a--top panel), $\Delta\Gamma$ decreases from zero without a sign change as $H_\mrm{ext}$ is increased. On the other hand, for the chiral theory, $\Delta\Gamma>0$ for $H_\mrm{ext}=0$, due to
stronger coupling of $m_s=0\rightarrow-1$ to the magnon-generated fields (cf. Fig.~\ref{TheoryDiff}a--bottom panel), which, when combined with the decrease of $\Delta\Gamma$ with $H_\mrm{ext}$, imprints the predicted sign change.

\begin{figure}[hbtp]
\centering
\includegraphics[width=0.48\textwidth]{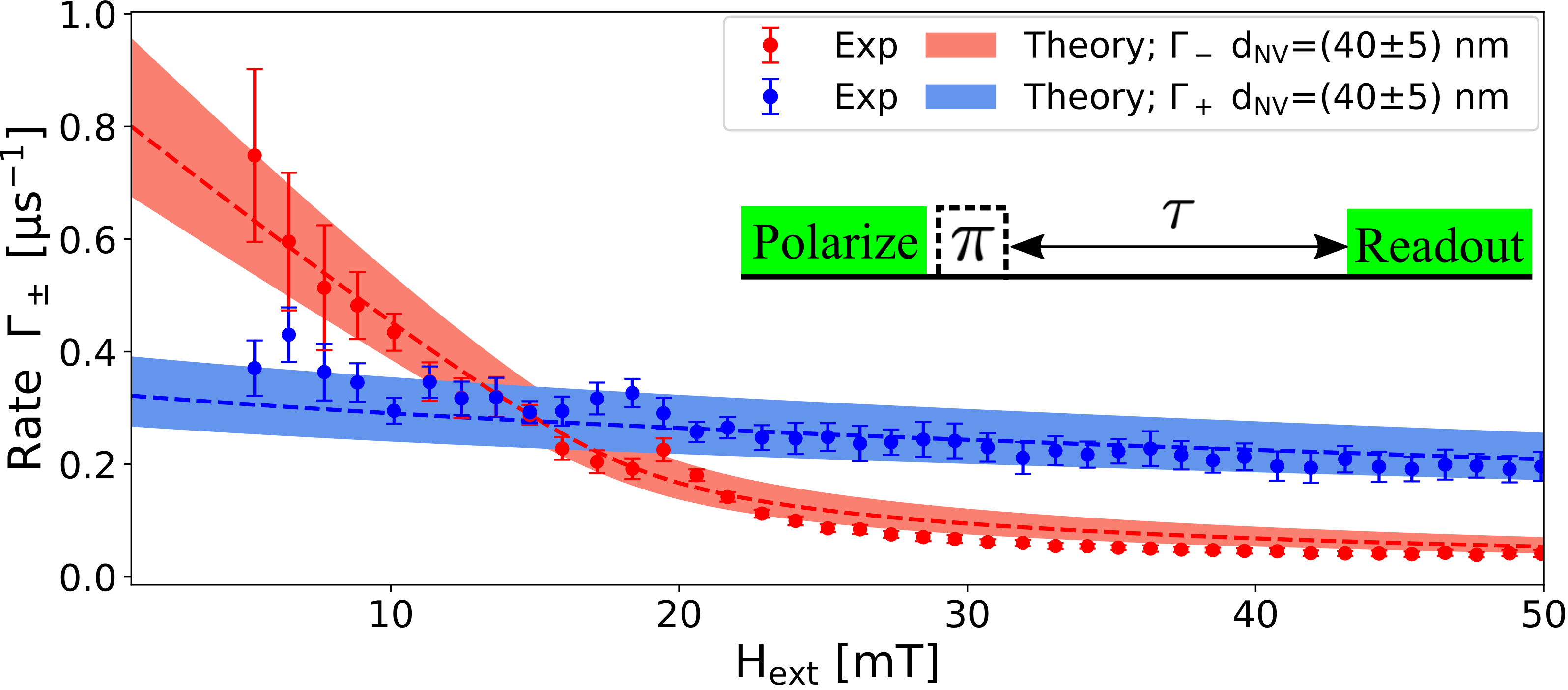}
\caption{\label{NickelPlot}Chiral coupling reflected in the crossover between transition rates $\Gamma_-$ and $\Gamma_-$ as a function of $H_\mrm{ext}$ in a NV-Nickel thin film hybrid. The measurement protocol is shown in the inset. Parameters for theoretical plots are the same as that mentioned in the caption of Fig.~\ref{TheoryDiff}.}
\end{figure}
\textit{Experiment}| To check the above picture experimentally, we next present our relaxometry measurements performed on NV-Nickel (Ni) thin film hybrids designed with the material and geometrical parameters needed to observe the sign change in $\Delta\Gamma$ due to chiral coupling-induced non-degenerate relaxations \cite{whyNickel}. We used a diamond with individually addressable NV centers implanted 10 nm below the surface, deposited a 30 nm SiO$_2$ spacer layer, and evaporated 40 nm of Ni on top. The two rates were determined by initializing the spin in each of its eigenstates (using a laser pulse and microwave pulses on the appropriate ESR transitions), waiting for a time $\tau$ that was swept, and characterizing the spin-dependent photoluminescence during a subsequent laser readout pulse (see inset of Fig.~\ref{NickelPlot}) \cite{van2015nanometre}. 
Fig.~\ref{NickelPlot} shows a crossover between the $\Gamma_+$ and  $\Gamma_-$ rates at $\sim 15$ mT which as seen from Fig.~\ref{TheoryDiff}b-c, is a clear signature that can only be captured by the presented ‘chiral’ theory for evaluating the QI spin relaxation rates.

\textit{Conclusions and outlook}| In summary, we show that the magnetic-dipole transitions hosted within typical spin$>$1/2 QIs couple disparately to the chiral magnetic noise 
produced by thermally populated magnons in nearby magnetic films. Via presenting an experimentally benchmarked theory, we demonstrate that including this, so far neglected, role of chirality is central for quantitatively and qualitatively understanding QI-relaxometry experiments. Our results will thus assist the recently growing effort of utilizing QI-relaxometry for understanding the dynamical properties of magnetic films to advance magnonics and quantum technologies utilizing magnets.

Our results can be extended to other phenomena generating chiral magnetic noise at the QI. For example, thermally populated one-way propagating magnetic modes at the boundaries of media with differing bulk band topology (as predicted by the recent theoretical proposal of topological magnon insulator \citep{shindou2013chiral,kim2017chiral})  provides a scenario where asymmetric magnon spectrum generates chiral magnetic noise. Such chiral modes also exist in electronic systems \citep{hasan2010rmp,tokura2019magnetic}, where electrical charge fluctuations in the mode would produce the chiral magnetic noise. Combining the advantages offered by QI-relaxometry (non-invasiveness, cryogenic to room temperature operation, up to nm-spatial and GHz frequency resolution) with the inherent chiral nature of QI-sensor highlighted here, may thus open new avenues for probing the above-mentioned phenomena of broad interest to the condensed matter community. 

\vspace{0.1 in}
AR, PU acknowledge support from the National Science Foundation through Grant No. DMR-1838513 and Grant No. ECCS-1810494.

\bibliography{refs}

\begin{thebibliography}{39}
\bibitem [{ell()}]{ellipticfield}%
  \BibitemOpen
  \href@noop {} {}\bibinfo {note} {We note that the dipolar field from
  propagating magnons are elliptical in nature due to the finite mode
  ellipticity of magnons. However, in the schematic in Fig.~\ref{Schematic}b-c,
  we show the dominant NV ESR transition.}\BibitemShut {Stop}%
\bibitem [{the()}]{theorydiff}%
  \BibitemOpen
  \href@noop {} {}\bibinfo {note} {To understand the behavior of $\Gamma_\pm$
  when scanning $\omega_\pm$, we can consider the thin film limit $kL \ll 1$,
  where the integrand in Eq.~(\ref{rate_exp}) scales as $k^2 e^{-2kd_{NV}}$
  with wavevector. For $\omega_\pm \approx \omega_\mathrm{FMR}(\equiv
  \omega_{sw}(\vec{k}=0))$, the relevant wavevectors are $\vec{k} \approx 0$,
  for which the integrand in Eq.~(\ref{rate_exp}) is zero. For $\omega_\pm \gg
  \omega_\mathrm{FMR}$, large wavevectors $k \gg d_{NV}^{-1}$ cause the
  exponential dependence to suppress the integrand and the relaxation rate is
  small. Thus, $\Gamma_\pm$ maximizes at some intermediate $\omega_\mrm{max
  \pm}$.}\BibitemShut {Stop}%
\end{thebibliography}
\end{document}